\documentclass[prd,aps,preprint,groupedaddress,superscriptaddress,nofootinbib]{revtex4}

\usepackage{comment}
\usepackage[utf8]{inputenc} 
\usepackage{graphicx,color,overpic,mathtools}
\usepackage{amsthm,amsmath,amssymb,hyperref,mathrsfs}
\usepackage{braket,bm,bbm,setspace}
\PassOptionsToPackage{normalem}{ulem}
\usepackage{ulem} 
\usepackage{physics}
\usepackage{float}
\usepackage[makeroom]{cancel}
\usepackage[english]{babel}
\usepackage{graphicx}
\graphicspath{ {images/} }
\addto\captionsspanish{}
\hypersetup{
    colorlinks=false,
    pdfborder={0 0 0},
}

\definecolor{rRGB}{RGB}{169.2, 155.0, 0.5}

\begin{document}

\vspace*{-2cm}
\hfill YITP-22-84
\vspace*{0.0cm}
\title{Constraints on thermalizing surfaces from infrared observations of supermassive black holes}

\author{Ra\'ul Carballo-Rubio}
\affiliation{CP3-Origins, University of Southern Denmark, Campusvej 55, DK-5230 Odense M, Denmark}
\affiliation{Florida Space Institute, University of Central Florida, 12354 Research Parkway, Partnership 1, 32826 Orlando, FL, USA}
\author{Francesco Di Filippo}
\affiliation{Center for Gravitational Physics and Quantum Information, Yukawa Institute for Theoretical Physics, Kyoto University, Kyoto 606-8502, Japan}
\author{Stefano Liberati}
\affiliation{SISSA - International School for Advanced Studies, Via Bonomea 265, 34136 Trieste, Italy}
\affiliation{IFPU, Trieste - Institute for Fundamental Physics of the Universe, Via Beirut 2, 34014 Trieste, Italy}
\affiliation{INFN Sezione di Trieste, Via Valerio 2, 34127 Trieste, Italy}
\author{Matt Visser}
\affiliation{
School of Mathematics and Statistics, Victoria University of Wellington, PO Box 600, Wellington 6140, New Zealand
}

\begin{abstract}

Infrared observations of Sgr A$^*$ and M87$^*$ are incompatible with the assumption that these sources have physical surfaces in thermal equilibrium with their accreting environments. In this paper we discuss a general parametrization of the energy balance in a horizonless object, which permits to quantify how close a horizonless object is in its behavior to a black hole, and analyze the timescale in which its surface can thermalize. We show that the thermalization timescale is unbounded, growing large for objects that mimic closely the behavior of a black hole (and being infinite for the latter). In particular, the thermalization timescale is proportional to the time that energy spends inside the horizonless object due to propagation and interactions with the bulk. Hence, these observations can be used to quantitatively restrict the dynamical behavior of horizonless objects, without being able to discard the existence of a physical surface.

\end{abstract}

\maketitle
\def\O{{\mathcal{O}}}
\newcommand{\red}[1]{{\color{red} #1}}
\newcommand{\blue}[1]{{\color{blue} #1}}

\section{Introduction}

While black holes have been for a long time a central topic in gravitation theory, the fast-pacing advancements in gravitational-wave detection and very-long-baseline interferometry (VLBI) observations have revived the interest in the possibility of probing the inner structure of these purely gravitational objects. Among the most striking consequences of these developments is the possibility to test deviations from the standard solutions of general relativity describing black holes, which are singular and are therefore expected to be regularized by quantum-gravitational effects.

Quite remarkably, the viable resulting geometries endowed with an outer horizon where found to belong to basically two families of solutions~\cite{Carballo-Rubio:2019fnb,Carballo-Rubio:2019nel}. Both these families admit as limiting cases horizonless ultracompact configurations (see~\cite{Carballo-Rubio:2022nuj} for details). Similar static solutions for ultra-compact quasi-black hole configurations can be found in the literature independently from the aforementioned limiting procedure (see e.g.~\cite{Laughlin:2003yh,Mazur:2004fk,Mazur:2001fv,Carballo-Rubio:2017tlh,Arrechea:2021pvg,Arrechea:2021xkp} and references therein), and as such they appears to be a rather generic class of black hole mimickers and interesting case study for observational constraints.  


While there is by now a rich literature concerning the theory, phenomenology and viable constraints in different classes of black hole mimickers (see e.g.~\cite{Carballo-Rubio:2018jzw,Cardoso:2019rvt} for comprehensive reviews on this subject), for what concerns constraints on ultra-compact horizonless objects with a physical surface, a special role has been recently played by VLBI observations of supermassive black holes (Sgr A$^*$ and M87$^*$)~\cite{EventHorizonTelescope:2019dse,EventHorizonTelescope:2022xqj}. Here, we will focus on complementary arguments that constrain the possible existence of a surface using infrared observations of Sgr A$^*$ and M87$^*$~\cite{Broderick:2005xa,Broderick:2007ek,Narayan:2008bv,Broderick:2009ph,Broderick:2015tda,EventHorizonTelescope:2022xqj}. 



The arguments in the aforementioned papers were groundbreaking in demonstrating that constraining the existence of a surface was within reach with available data from infrared observations. In particular, these papers indicate that observations are incompatible with a physical surface in thermal equilibrium with its environment. Nonetheless, our understanding of black hole mimickers has strongly advanced in recent times, and it is not clear whether thermal equilibrium is reached within a sufficiently short timescale. In what follows we shall show that a more accurate characterization of the physics involved in these exotic objects has a profound impact on the implications of these early analyses, resulting in more complete physical models and thus refined constraints.


The present authors have pursued this line of research in previous works, in particular~\cite{Carballo-Rubio:2018jzw} and~\cite{Carballo-Rubio:2022imz} (see also~\cite{Lu:2017vdx,Cardoso:2017cqb} by other authors). These works have shown that updating the assumptions in~\cite{Broderick:2005xa,Broderick:2007ek,Narayan:2008bv,Broderick:2009ph,Broderick:2015tda} can result in sizeable changes in the associated constraints, thus reaffirming the necessity for a critical revision of the underlying assumptions on which the latter are based.

We want to stress here that the most critical aspect for the evaluation of these constraints is an adequate parametrization of the energy exchange between the horizonless object and its environment. More specifically, equilibrium requires that incident energy onto the horizonless object is re-emitted, which will generally occur only after a certain re-emission timescale. It is essential to account for this re-emission timescale in analysis that determine whether or not reaching equilibrium is possible. In this work, we study this problem for the first time, building a general parametrization of this energy exchange that includes a temporary absorption coefficient and timescale, and analyze how the equilibrium timescale depends on these parameters. 


\section{Energy balance in a horizonless object \label{sec:channels}}

When a black hole is surrounded by matter, all energy that moves across the horizon is absorbed by the black hole, which adjusts dynamically by changing its mass and angular momentum (and possibly electric charge, though this is not particularly relevant in astrophysical situations). Of course, semiclassically black holes can in principle re-emit part of this energy back in the form of Hawking radiation over long times, however for most astrophysical black holes the cosmic microwave background is hot enough to counterbalance this tendency and induces further black hole growth (measured in terms of the horizon area) even in the absence of matter fluxes~\cite{Taylor:2022pgs}. 

For horizonless objects the physics is more complex. In the most general situation, the net absorption associated with a black hole can be replaced by the following channels:
\begin{enumerate}
    \item \textbf{Absorption:} A fraction $\kappa$ of the incident energy can be permanently absorbed by the internal degrees of freedom of the object, changing the intrinsic state of the latter.
    \item \textbf{Temporary absorption/Delayed re-emission:} A fraction $\tilde{\kappa}$ of the incident energy can be re-emitted (inelastically) after a certain amount of time $\tau_{\tilde{\kappa}}$, with the delay caused by a combination of propagation and interaction effects in the bulk.
    \item \textbf{Instantaneous re-emission:} A fraction $\tilde{\Gamma}$ of incident energy can be re-emitted (inelastically) almost instantaneously, after interaction with surface degrees of freedom.
    \item \textbf{Reflection:} A fraction $\Gamma$ of incident energy can be reflected (elastically) without being absorbed by the object.
    \item \textbf{Transmission:} A fraction $\mathrm{T}$ of the incident energy can travel freely across the object without any interactions taking place.
\end{enumerate}
Conservation of energy implies $\kappa+\tilde{\kappa}+\tilde{\Gamma}+\Gamma + \mathrm{T}=1$.

Note that the coefficient $\tilde{\kappa}$ can either describe absorption or instantaneous re-emission in the limits $\tau_{\tilde{\kappa}}\rightarrow \infty$ and $\tau_{\tilde{\kappa}}\rightarrow 0$, respectively. Hence, it can be understood as a more physical realization of these two (idealized) channels. In previous work
~\cite{Carballo-Rubio:2018jzw}, when applying this parametrization to a discrete model of energy exchange, we only considered these idealized channels (also, we implicitly set $\mathrm{T}\to0$), but here we want to go a step forward.

The specific behavior of a given horizonless object is model-dependent. In fact, our knowledge of the dynamics of these objects is not detailed enough to determine which of the channels above is dominant for a given model. Hence, from a phenomenological perspective it is reasonable to consider all of them as equally possible, and cast constraints on the different parameters involved.

Given the above five parameters $\kappa$, $\tilde{\kappa}$, $\tilde{\Gamma}$, $\Gamma$ and $\mathrm{T}$, respectively introduced for the five items listed above, we can easily see that they are sufficient for characterizing a broad class of horizonless black hole mimickers. One of such objects with only $\kappa\neq0$ will be the closest in behavior to a black hole. On the other hand, a horizonless object with only $\tilde{\kappa}\neq0$ will behave like a black hole for a certain timescale $\tau_{\tilde{\kappa}}$ that can be very long depending on the model. The remaining limiting cases, only $\tilde{\Gamma}\neq0$ and only ${\Gamma}\neq0$ respectively, display more stark deviations with respect to black holes, and could be potentially constrained in VLBI observations~\cite{EventHorizonTelescope:2022xqj,Carballo-Rubio:2022aed}. A similar comment applies to objects with only $\mathrm{T}\neq0$~\cite{Eichhorn:2022oma,Eichhorn:2022fcl}.

Now that we have introduced our parametrization, let us dwell in the next section in the discuss previous works that have explored the role of these parameters in infrared and VLBI observations of supermassive black holes.

\section{Relation to previous work}

The parametrization introduced in the previous section aimed at being complete regarding the possible types of interactions between the incoming energy and the horizonless object. Previous works in the subject consider a subset of these behaviors, which we briefly review in the following, together with the reasons behind such choices.

Most of the works below assumed spherical symmetry (except when otherwise noted below). Hence, we can introduce an effective radius of the object $R$, together with a dimensionless measure of compactness, $\mu=(R-2M)/2M$. 

\begin{itemize}
    \item The original works~\cite{Broderick:2005xa,Broderick:2007ek,Narayan:2008bv,Broderick:2009ph,Broderick:2015tda} assumed an instantaneous remission by the ultra compact object of the incident radiation, i.e.~$\kappa=\tilde{\kappa}=\Gamma=\mathrm{T}=0$, and only $\tilde{\Gamma}\neq0$. In the argument provided by the authors this follows from the consideration that, in thermal equilibrium, Kirchhoff's law implies that all energy received by the horizonless object is instantly re-emitted. On general grounds this would imply that, if energy is initially distributed among the other channels for a given model of horizonless object, it is the dynamical evolution towards equilibrium that progressively re-distributes it until the energy balance can be adequately described by $\kappa=\tilde{\kappa}=\Gamma=\mathrm{T}=0$. The authors showed then that thermal equilibrium is incompatible with infrared observations.
    \item A further step was taken in~\cite{Cardoso:2017njb,Lu:2017vdx} with the analysis of the timescale required for equilibrium to be reached, still under the assumption $\kappa=\tilde{\kappa}=\Gamma=\mathrm{T}=0$ so that only $\tilde{\Gamma}\neq0$. This analysis showed that gravitational lensing plays an important role in attaining thermal equilibrium, a role previously unaccounted for. Indeed, with increasing compactness there is a closing escaping angle $\Delta \Omega$ for rays leaving the object surface: for $\mu\ll 1$ one can show that $\Delta\Omega/2\pi \approx 27\mu/8 +O(\mu^2)$~\cite{Carballo-Rubio:2018jzw} (in the following, we will define $\Delta=\Delta\Omega/2\pi$). In turn this implies that the timescale in which equilibrium is reached must scale at least as $1/\mu$. Thus, the equilibrium assumption fails to hold for $\mu$ small enough. This means that the incompatibility between thermal equilibrium and infrared observations can be translated into a constraint on $\mu$ (or, equivalently, $R$).
    \item The timescale to reach equilibrium was re-analyzed in~\cite{Carballo-Rubio:2018jzw}, together with the introduction of non-zero coefficients $\kappa\neq0$, $\Gamma\neq0$ and $\tilde{\Gamma}\neq0$ (the coefficient $\tilde{\kappa}$ was implicitly considered in the general parametrization introduced, but put to zero for the analysis of equilibrium, together with $\mathrm{T}=0$). For this more general situation, the constraints are now formulated as (generically nonlinear) combinations of the available parameters. Of particular importance is the absorption coefficient, being the constraints very sensitive to non-zero values of the latter.
    \item{Once rotation is included~\cite{Zulianello:2020cmx}, re-emission is not uniform throughout the surface. This effect increases with spin, and makes previous calculations of the timescale in which equilibrium is reached inapplicable. In particular, the re-emission pattern of equilibrium in the presence of rotation, and the timescale in which this pattern can arise, are unknown.}
    \item{In~\cite{EventHorizonTelescope:2022xqj}, an updated account of the original works~\cite{Broderick:2005xa,Broderick:2007ek,Narayan:2008bv,Broderick:2009ph,Broderick:2015tda} is provided, also taking into account the aforementioned effect of gravitational lensing~\cite{Cardoso:2017njb,Lu:2017vdx}. This updated discussion still has $\kappa=\tilde{\kappa}=\Gamma=\mathrm{T}=0$ and only $\tilde{\Gamma}\neq0$, as it focuses on the equilibrium state. That neglecting absorption is questionable was stressed in the follow-up paper~\cite{Carballo-Rubio:2022imz} stressing again the profound impact that taking it into account can have on the obtainable constraints.}
\end{itemize}

As it is apparent in the brief review above, the arguments~\cite{Broderick:2005xa,Broderick:2007ek,Narayan:2008bv,Broderick:2009ph,Broderick:2015tda} have generated widespread interest, and further refinements have been published by different groups. A possible point of contention is whether or not a non-zero value of $\kappa$ is physically reasonable. Let us discuss this in some detail in the next section.

For completeness, before focusing on the role of $\kappa$ and $\tilde{\kappa}$, we include a list of works that have used (part of) the parametrization above to model VLBI observations of alternatives to black holes. VLBI observations provide complementary constraints to the infrared constraints that are the subject of this paper. The parametrization introduced in~\cite{Carballo-Rubio:2018jzw}, which does not include the transmission coefficient, was used in~\cite{Carballo-Rubio:2022aed} to determine the image features associated with reflection and re-emission, providing an exhaustive exploration of the parameter space spanned by $R$, $\Gamma$ and $\tilde{\Gamma}$. Complementary models in which only non-zero transmission coefficient $\mathrm{T}$ was included were the focus of~\cite{Eichhorn:2022oma,Eichhorn:2022fcl}. On the other hand, ~\cite{EventHorizonTelescope:2022xqj} also discussed the features associated with reflection for specific values of the parameters $R$ and $\Gamma$.

\section{The interplay between absorption and thermal equilibrium}

It is clear that $\kappa\neq0$ prevents equilibrium, in the sense of perfect balance between received and instantaneoulsy re-emitted energy, to be reached. The same comment holds true for any of the other channels (delayed re-emission, reflection and transmission) discussed in Sec.~\ref{sec:channels}. Indeed, any energy deposited in any of these channels cannot go into the instantaneous re-emission channel, thus always resulting into a deficit in the re-emission channel with respect to the incident energy.

Hence, a possible objection is that assuming that $\kappa\neq0$ is incompatible with equilibrium. Note, however, that this is actually what happens if the central object is a classical black hole. A classical black hole can never be in equilibrium with its accreting environment, due to its purely absorptive nature ($\kappa=1$), and the fact that all incident energy is stored in internal degrees of freedom. The same comment applies to semiclassical black holes ($|\kappa-1|\ll 1$), as the features of re-emission of energy in the form of Hawking radiation are constrained as a function of the black hole mass, and cannot be arbitrarily adjusted to achieve equilibrium with the accreting environment.

Horizonless objects are expected to mimic closely the behavior of black holes. Even though the mimicked behaviors are model-dependent, it is reasonable to expect that at least part of the incident energy will be transferred to internal degrees of freedom, and that not all this energy can be re-emitted in arbitrary amounts to achieve equilibrium. A simple argument in this sense consisting in the fact that ultra-compact object must be able to convert at least some of the incident energy in expansion so to avoid to form a trapping horizon as a consequence of the accreting energy~\cite{Carballo-Rubio:2018vin}. 

These aspects are certainly dependent on the dynamics of specific models, which is not well understood. It is therefore completely unknown whether it is reasonable to assume that a horizonless object must reach equilibrium with its accreting environment. It may well be that not being able to reach equilibrium with its accreting environment (or not being able to do so on astrophysical relevant timescales) is a feature of horizonless objects.

Even if we assume that $\kappa=0$, there is still the issue that, while all incident energy in this case will be radiated away, the amount of time it takes for the re-emission to take place, $\tau_{\tilde{\kappa}}$, is unknown and also dependent on model-dependent dynamics. Again, for a classical black hole this time is infinite, so one can expect that a good black hole mimicker will have a relatively long timescale for delayed re-emission. This leads to the natural question about how this delay in re-emission impacts on the achievement of thermal equilibrium, in particular on the associated timescale.

As the role of absorption $\kappa$ has been studied in previous papers, we will focus on the role of temporary absorption for the rest of the paper. For the sake of comprehensiveness, we will discuss the energy exchange between a horizonles objects and its accreting environment in full generality, and then focus on the situation in which $\kappa=\tilde{\Gamma}=\Gamma=\mathrm{T}=0$ but $\tilde{\kappa}\neq0$, and analyze amount of time that it takes for a horizonless object to achieve equilibrium with its environment as a function of the timescale of energy release. We will discuss how this model reproduces the behavior analyzed previously in suitable limits (very short and very long re-emission times, respectively), and the new insights that it provides into the problem of equilibrium.


 \section{A discrete model of energy exchange} \label{sec:dismod}

In this section, we introduce a discrete model to describe the energy exchange between a general horizonless object and its environment.

Let us consider a discretization of time such that we use the set of integers $\{1,...,n\}$ to denote different moments in time. All time intervals have the same size $\Delta t$, which we take to be roughly proportional to the light-crossing time $\tau_{\rm S}=r_{\rm S}/c$. We assume that there is a uniform energy injection $x$ in each interval. Also, $\{x_i\}_{i=1}^n$ will be the incident energy (that is, the energy that reaches the object from its environment) at different moments, and $\{\epsilon_i\}_{i=1}^n$ the energy released by the object at the same time. 

\begin{figure}
    \centering
    \includegraphics{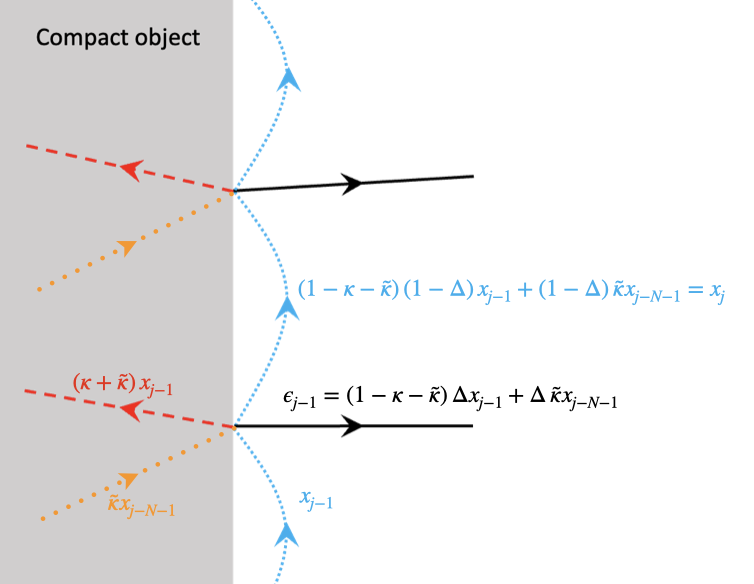}
    \caption{Schematic proof of Eqs.~\eqref{eq:rec_rel_eps} and~\eqref{eq:rec_rel_x}, with time in the vertical direction. The quantities $\epsilon_k$ and $x_k$ are the released energy and the incident energy in the interval $k$, respectively. These quantities can be related to each other and also to the corresponding quantities in the previous time interval $k-1$, as shown in the figure (see also App.~\ref{sec:app} for a complementary discussion).}
    \label{fig:my_label}
\end{figure}

In App.~\ref{sec:app} we discuss the energy balance for different time intervals in order to derive recursion relations, while Fig.~\ref{fig:my_label} provides a schematic summary. It follows that we can write the total incident energy $X_n$ and the total escaping energy $E_n$ in the interval $n\leq N$ as
\begin{equation}\label{eq:rr1}
X_n=\sum_{k=1}^nx_k,\qquad E_n=\sum_{k=1}^n\epsilon_k,
\end{equation}
where
\begin{equation}\label{eq:rr2}
\epsilon_k=\Delta\tilde{\Gamma}x_k,\qquad 1\leq k\leq N,
\end{equation}
while
\begin{equation}\label{eq:rr3}
x_1=x,\qquad x_2=(1-\Delta)\tilde{\Gamma}x_1,
\end{equation}
and
\begin{equation}\label{eq:rr4}
x_{k+1}=[\Gamma+(1-\Delta)\tilde{\Gamma}]x_k,\qquad 2\leq k\leq N.   
\end{equation}
Due to temporary absorption, Eqs.~\eqref{eq:rr1}-\eqref{eq:rr4} must be completed with the following modifications for $n\geq N+1$:
\begin{equation}
\epsilon_{k}=\Delta(\tilde{\Gamma}x_{k}+\tilde{\kappa}x_{k-N}),\qquad k\geq N+1, 
\end{equation}
and
\begin{equation}
x_{k}=[\Gamma+(1-\Delta)\tilde{\Gamma}]x_{k-1}+(1-\Delta)\tilde{\kappa}x_{k-N-1}\qquad k\geq N+1.
\end{equation}
Let us now take a closer look to the physics in these recursion relations.

\section{Role of temporary absorption}

The recursion relations discussed in the previous section allow to study general situations for arbitrary values of the five parameters $\kappa$, $\tilde{\kappa}$, $\tilde{\Gamma}$, $\Gamma$ and $\mathrm{T}$. However, as we want to understand the role played by temporary absorption in the achievement of thermal equilibrium, we will focus here on the simplified case in which only $\kappa$ and $\tilde{\kappa}$ are non-zero. The recursion relations are then reduced to
\begin{equation}\label{eq:rec_rel_eps}
\epsilon_{k}=\Delta[(1-\kappa-\tilde{\kappa})x_{k}+\tilde{\kappa}x_{k-N}],\qquad k\geq N+1, 
\end{equation}
and
\begin{equation}\label{eq:rec_rel_x}
x_{k}=(1-\Delta)(1-\kappa-\tilde{\kappa})x_{k-1}+(1-\Delta)\tilde{\kappa}x_{k-N-1}\qquad k\geq N+1.
\end{equation}
These expressions cannot be summed analytically. Numerical evaluation is always possible, though we have also been able to find an analytical approximation as discussed in the following.

For the purposes of finding a suitable analytical approximation, let us consider for a moment $\tilde{\kappa}=0$ (no temporary absorption), for which the recursion relation can be summed analytically leading to~\cite{Carballo-Rubio:2018jzw,Carballo-Rubio:2022imz}:
\begin{equation}\label{eq:flux_noreemission}
       \frac{\dot E\left(\tilde\kappa=0\right)}{\dot M}=\frac{\Delta(1-\kappa)}{\kappa+\Delta(1-\kappa)}\left\{ 1-(1-\kappa)^{t/\tau_S}\left(1-\Delta\right))^{t/\tau_S}	\right\}\,. 
\end{equation}
From this expressions, it follows that for timescales longer than
\begin{equation}
    t\gtrsim \min\left(\kappa^{-1},\Delta^{-1}\right)\tau_S,
\end{equation}
the outgoing flux reaches a steady state:
\begin{equation}\label{eq:flux_steady}
    \left. \frac{\dot E\left(\tilde\kappa=0\right)}{\dot M}\right|_{\rm steady}\simeq\frac{\Delta(1-\kappa)}{\kappa+\Delta(1-\kappa)}\,.
\end{equation}

\begin{figure}
    \centering
    \includegraphics[width=.49\linewidth]{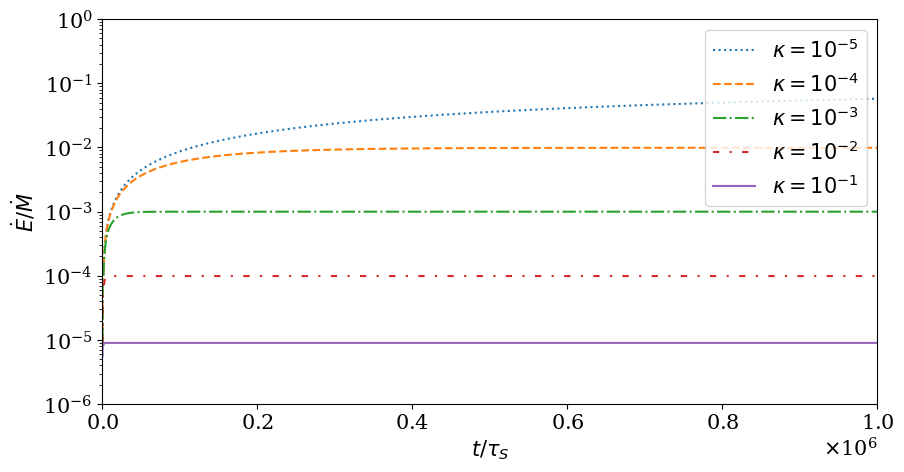}
    \includegraphics[width=.49\linewidth]{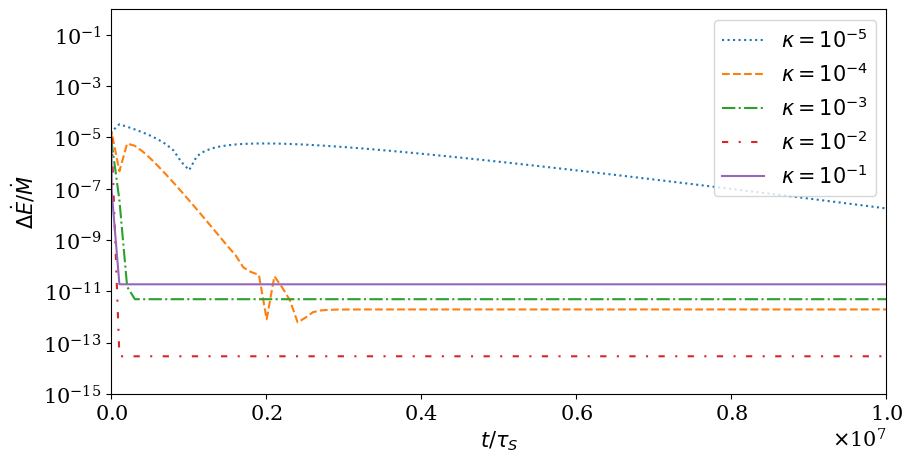}       
    \includegraphics[width=.49\linewidth]{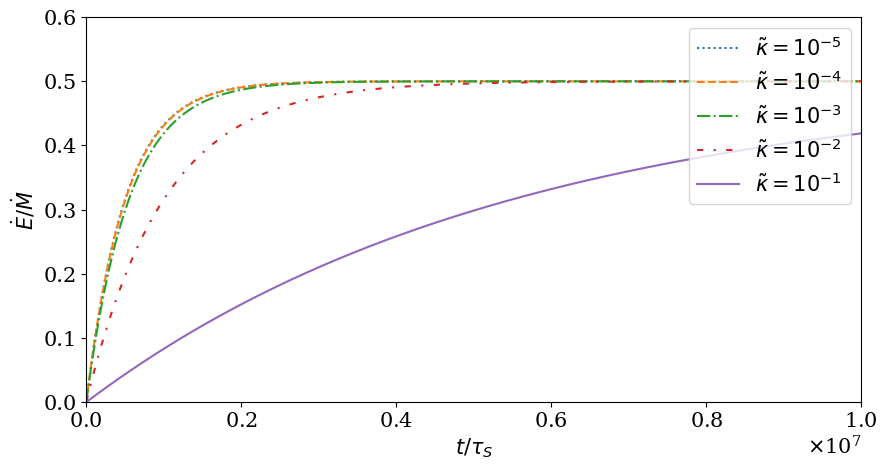}
    \includegraphics[width=.49\linewidth]{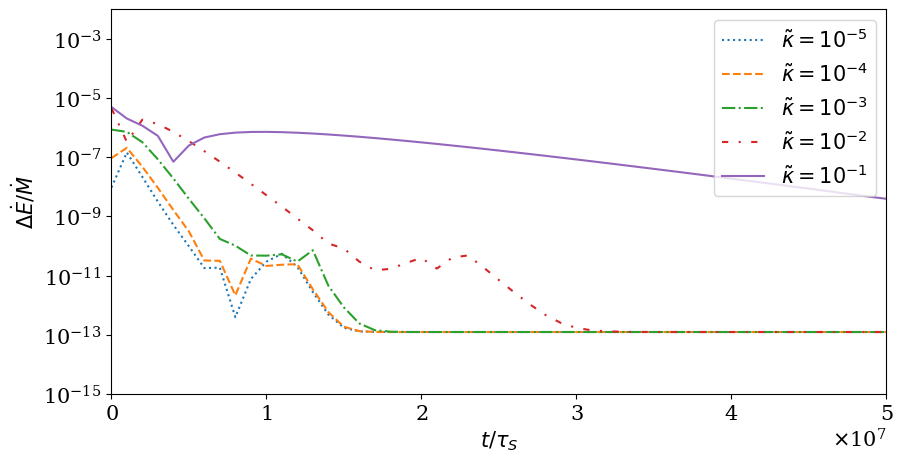}    
    \caption{The left panels show the numerical evaluation of the outgoing flux of energy, while the right panels show the difference $\Delta\dot E/M$ between the numerical evaluation and the analytical guess given in Eq.~\eqref{eq:guess}. The top panels are obtained fixing  $\Delta=10^{-6}$, $\tilde\kappa=10^-2$, $N=10^3$ and varying  $\kappa$. The bottom panels are obtained fixing  $\Delta=10^{-6}$, $\kappa=10^{-6}$, $N=10^2$ and varying  $\tilde\kappa$. }\label{fig:analytic_numeric}
\end{figure}

Let us now come back to the case in which $\tilde\kappa\neq 0$. Delayed re-emission introduces a delay between two successive bounces on the surface for a fraction of the energy. It is then reasonable to conjecture that considering the expression without re-emission, and replacing the timescale $\tau_\text{S}$ with the average timescale between to consecutive bounces for the fraction of energy that eventually escapes the gravitational field, could provide a good analytical approximation. A fraction $\tilde\kappa$ of the energy takes a time $\tau_S+\tau_{\tilde{\kappa}}=(N+1)\tau_{\text{S}}$ between two consecutive bounces, whereas the  remaining energy takes a time $\tau_{\text{S}}$. Therefore, the average time is given by
\begin{equation}
\bar{\tau}=\tilde\kappa(N+1)\tau_{\text{S}} + (1-\tilde \kappa)\tau_{\text{S}}=(\tilde \kappa N+1)\tau_{\text{S}}\,,
\end{equation}
and we can make the following analytical guess:
\begin{equation} \label{eq:guess}
    \frac{\dot E_\text{guess}}{\dot M}=\frac{\Delta(1-\kappa)}{\kappa+\Delta(1-\kappa)}\left\{ 1-(1-\kappa)^{t/\bar\tau}\left(1-\Delta\right)^{t/\bar\tau}	\right\}.
\end{equation}
It is straightforward to check numerically whether this provides a good approximation for the flux of energy; Fig.~\ref{fig:analytic_numeric} shows that this is indeed the case. We can therefore use Eq.~\eqref{eq:guess} as a very good approximation of the outgoing flux of energy. From this result, we can infer that the presence of delayed re-emission does not alter the asymptotic value of the energy flux once the steady state is achieved; rather, it prolongs the time it takes to reach the steady state. In fact, the steady state is now reached for timescales
\begin{equation}
    t\gtrsim \min\left(\kappa^{-1},\Delta^{-1}\right)(\tilde \kappa N+1)\tau_{\text{S}}
\end{equation}

We can see that $N$ plays an important role in the thermalization timescale. While $\tilde{\kappa}$ is by construction bounded from above by 1, $N$ can be unbounded. In fact, taking the limit $N\rightarrow \infty$ recovers the behavior of a black hole, which means that larger values of $N$ yield better black hole mimickers.
In fact, even in the absence of absorption, the presence of temporary absorption can significantly weaken the constraint. For instance, in the case of Sgr A*, if we assume $\kappa = \tilde\kappa = 0$ the observational constraint~\cite{EventHorizonTelescope:2022xqj}
\begin{equation}
    \frac{\dot E}{\dot M}<10^{-3}\,,
\end{equation}
(note that~\cite{EventHorizonTelescope:2022xqj} provides a tighter constraint of $10^{-3}$ instead of the $10^{-2}$ in the original paper~\cite{Broderick:2005xa}) implies
\begin{equation}\label{eq:const_no_reemis}
    \frac{\dot E}{\dot M}\simeq \mu \frac{T}{\tau_S}<10^{-3}\,,
\end{equation}
where we have used the Eddington timescale $T\simeq 3.8\times 10^8\mbox{ yr}$ to provide an estimation of the typical timescale for the variation of its accretion rate. Note that the argument above requires the stationary of the source to be strictly applicable. Source variability can disrupt equilibrium or delay its onset in a way that is difficult to estimate using the formalism above. In some cases (e.g.~\cite{Cardoso:2017njb}), the Hubble time is used instead of the Eddington timescale, which changes the bounds below but can also lead to an overestimation of these constraints due to the non-equilibrium nature of the source. Another aspect to take into account is that the accretion rate used in the equations above is also changing in time and likely higher in the past. Hence, there is some ambiguity on the precise numerical values of these constraints; a definitive solution for these ambiguities would require a more thorough understanding of the evolution of the coupled system composed by the horizonless central object and its accreting envinroment.

Equation \eqref{eq:const_no_reemis} implies
\begin{equation}
    \mu<10^{-18}\,.
\end{equation}
On the other hand, when $\tilde\kappa\neq 0$, we get
\begin{equation}
    \mu<\frac{T}{(\tilde \kappa N+1)\tau_{\text{S}}}\,,
\end{equation}
This constraint can be much weaker than the one given in  Eq.~\eqref{eq:const_no_reemis} for $N$ large enough.
A fundamental question to answer is therefore the value that $N$ typically takes for specific models such as gravastars~\cite{Laughlin:2003yh,Mazur:2004fk,Mazur:2001fv} or semiclassical relativistic stars~\cite{Carballo-Rubio:2017tlh,Arrechea:2021pvg,Arrechea:2021xkp}. Unfortunately, the dynamics of these models is not yet understood well enough to extract the value of $N$. Nevertheless, it is possible to illustrate that $N$ can become very large for black hole mimickers, due to gravitational time delay associated with propagation effects. 

Let us consider a very simple toy model, constructed in spherical symmetry by demanding that the Misner--Sharp--Hernandez mass~\cite{Misner:1964je,Hernandez:1966zia} for each sphere is an $\epsilon$ away from its critical value which would yield the formation of a horizon~\cite{Nielsen:2008kd}. The interior of such a stellar structure~\cite{Guendelman:1991qb,Guendelman:1993qf,Brustein:2023cvf} is approximately described by the metric
\begin{equation}~\label{eq:eps_metric}
\text{d}s^2=-\epsilon\text{d}t^2+\frac{1}{\epsilon}\text{d}r^2+r^2\text{d}\Omega^2,
\end{equation}
where $\text{d}\Omega^2$ is the usual line element on the unit 2-sphere. The re-emission timescale for incident energy can be split as the sum of the time of propagation inside the structure, plus the interaction with the latter. From Eq.~\eqref{eq:eps_metric}, we can see that just propagation effects imply for this model that
\begin{equation}
N\gtrsim\frac{1}{\epsilon}.
\end{equation}
For $\epsilon\ll 1$, we then have
\begin{equation}
\bar{\tau}\sim \frac{\tilde{\kappa}}{\epsilon}\tau_{\rm S}\simeq \frac{10^{-22}\tilde{\kappa}}{\epsilon}\left(\frac{M}{M_\odot}\right)\tau_{\rm H},
\end{equation}
where $\tau_{\rm H}$ is the Hubble time and $M_\odot$ the mass of the Sun. Hence, it is not difficult to have ultracompact objects for which the thermalization timescale becomes comparable or even larger than the Hubble time, which means that thermalization is not possible for these objects in practice if $\epsilon\lesssim 10^{-22}(M/M_\odot)\tilde{\kappa}$. Let us stress that using the Hubble time is a conservative estimate, and a more realistic estimate would be provided by the variability timescale for a particular astrophysical system (e.g. Sgr A$^*$), which should be several orders of magnitude lower than $\tau_{\rm H}$.

\section{Conclusions}

Modeling the interactions between horizonless objects and their accreting environments is essential to cast constraints on these alternatives to black holes. In this paper, we have presented a general parametrization of these interactions, and focused on understanding the role that temporary absorption plays in reaching steady state.

Temporary absorption is necessary for the horizonless object to be adapt dynamically to its environment and be eventually be able to reach a steady state. This is in particular necessary to avoid the formation of horizons. Hence, a non-zero value of $\tilde{\kappa}$ seems to be unavoidable based on known physics.

The second parameter necessary to describe temporary absorption is the re-emission timescale, which we have parametrized in terms of $N$. We have shown that this parameter has an important impact on the thermalization timescale and that it can become arbitrarily large, preventing the latter from happening altogether for relatively compact horizonless objects.

In summary, that equilibrium is not observed in systems such as Sgr A$^*$ and M87$^*$ can be used to place constraints on horizonless objects, ruling out models in which this thermalization timescale is short enough so that expecting equilibrium is reasonable. However, we have shown that simple arguments indicate that ultracompact objects would have thermalization timescales that are too long for equilibrium to be feasible in our universe. Hence, it is possible that supermassive horizonless objects, not in equilibrium with their accreting environments, exist in nature.

\appendix

\section{Discretized energy exchange} \label{sec:app}

Let us discuss in detail the different components at play in the energy exchange between a horizonless objects and its environment, within the discretized model used in the paper. This provides a derivation of the recursion relations in Sec.~\ref{sec:dismod}.

For the first interval, the energy balance is as follows:
\begin{itemize}
\item There is an injection of energy $x$ onto the horizonless object. 
\item The total amount of incident energy is $x_1=x$.
\item A fraction of energy $\kappa x_1$ is permanently absorbed by the object.
\item A fraction $\tilde{\kappa}x_1$ is temporarily absorbed by the object, and will be re-emitted after a time $\tau_{\tilde{\kappa}}=N\tau_{\rm S}$.
\item A fraction $\tilde{\Gamma}x_1$ is re-emitted instantaneously. From this fraction, an amount $\Delta\tilde{\Gamma} x_1$, where $\Delta=\Delta\Omega/2\pi$, escapes the gravitational well of the object, while the remaining amount $(1-\Delta)\tilde{\Gamma}x_1$ is gravitationally lensed back to the object. Let us define $\epsilon_1=\Delta\tilde{\Gamma} x_1$.
\item A fraction $\Gamma x_1$ is reflected, escaping the gravitational well of the object.
\item A fraction $\mathrm{T} x_1$ travels across the object without interaction.
\end{itemize}
For the second interval:
\begin{itemize}
\item There is an injection of energy $x$ onto the horizonless object. 

\item A fraction of energy $(1-\Delta)\tilde{\Gamma}x_1$ returns to the surface after being re-emitted in the first interval.

\item The total amount of incident energy is $x_1+x_2$, where $x_2=(1-\Delta)\tilde{\Gamma}x_1$.

\item A fraction of energy $\kappa(x_1+x_2)$ is permanently absorbed by the object.

\item A fraction $\tilde{\kappa}(x_1+x_2)$ is temporarily absorbed by the object, and will be re-emitted after a time $N\tau_{\rm S}$.
\item A fraction $\tilde{\Gamma}(x_1+x_2)$ is re-emitted instantaneously. From this fraction, an amount $\Delta\tilde{\Gamma}(x_1+x_2)$ escapes the gravitational well of the object, while the remaining amount $(1-\Delta)\tilde{\Gamma}(x_1+x_2)$ is gravitationally lensed back to the object. Let us define $\epsilon_2=\Delta\tilde{\Gamma}x_2=(1-\Delta)\tilde{\Gamma}\epsilon_1$, so that the total energy that escapes is $\epsilon_1+\epsilon_2$.
\item A fraction $\Gamma(x_1+x_2)$ is reflected. From this fraction, an amount $\Gamma x_1$ escapes the gravitational well of the object, while the remaining amount $\Gamma x_2$ is gravitationally lensed back to the object.
\item A fraction $\mathrm{T}(x_1+x_2)$ travels across the object without interaction.
\end{itemize}
For the third interval:
\begin{itemize}
\item There is an injection of energy $x$ onto the horizonless object. 

\item A fraction of energy $(1-\Delta)\tilde{\Gamma}(x_1+x_2)=x_2+(1-\Delta)\tilde{\Gamma}x_2$ returns to the surface after being re-emitted in the previous interval.

\item A fraction of energy $\Gamma x_2$ returns to the surface after being reflected in the previous interval.

\item The total amount of incident energy is $x_1+x_2+x_3$, where $x_3=[\Gamma+(1-\Delta)\tilde{\Gamma}]x_2$.

\item A fraction of energy $\kappa(x_1+x_2+x_3)$ is permanently absorbed by the object.

\item A fraction $\tilde{\kappa}(x_1+x_2+x_3)$ is temporarily absorbed by the object, and will be re-emitted after a time $N\tau_{\rm S}$.
\item A fraction $\tilde{\Gamma}(x_1+x_2+x_3)$ is re-emitted instantaneously. From this fraction, an amount $\Delta\tilde{\Gamma}(x_1+x_2+x_3)$ escapes the gravitational well of the object, while the remaining amount $(1-\Delta)\tilde{\Gamma}(x_1+x_2+x_3)$ is gravitationally lensed back to the object. Let us define $\epsilon_3=\Delta\tilde{\Gamma}x_3=(1-\Delta)\tilde{\Gamma}\epsilon_2$, so that the total energy that escapes is $\epsilon_1+\epsilon_2+\epsilon_3$.

\item A fraction $\Gamma(x_1+x_2+x_3)$ is reflected. From this fraction, an amount $\Gamma x_1$ escapes the gravitational well of the object, while the remaining amount $\Gamma(x_2+x_3)$ is gravitationally lensed back to the object.
\item A fraction $\mathrm{T}(x_1+x_2+x_3)$ travels across the object without interaction.
\end{itemize}

For the $(N+1)-$interval:
\begin{itemize}
\item There is an injection of energy $x$ onto the horizonless object. 

\item A fraction of energy $(1-\Delta)\tilde{\Gamma}\sum_{k=1}^{N}x_n$ returns to the surface after being re-emitted in the previous interval.

\item A fraction of energy $\Gamma\sum_{k=2}^{N}x_n$ returns to the surface after being reflected in the previous interval.

\item The total amount of incident energy is $X_{N+1}=\sum_{k=1}^{N+1}x_k$, where $x_{k}=[\Gamma+(1-\Delta)\tilde{\Gamma}]x_{k-1}$.

\item A fraction of energy $\kappa X_{N+1}$ is permanently absorbed by the object.

\item A fraction $\tilde{\kappa}X_{N+1}$ is temporarily absorbed by the object, and will be re-emitted after a time $N\tau_{\rm S}$.

\item A fraction $\tilde{\kappa}x_1$ is re-emitted after being temporarily absorbed by the object, while a fraction $\tilde{\Gamma}X_{N+1}$ is re-emitted instantaneously. From these fractions, an amount $\Delta(\tilde{\Gamma}X_{N+1}+\tilde{\kappa}x_1)$ escapes the gravitational well of the object, while the remaining amount $(1-\Delta)(\tilde{\Gamma}X_{N+1}+\tilde{\kappa}x_1)$ is gravitationally lensed back to the object. Let us define $\epsilon_{N+1}=\Delta(\tilde{\Gamma}x_{N+1}+\tilde{\kappa}x_1)$, so that the total energy that escapes is $E_{N+1}=\sum_{k=1}^{N+1}\epsilon_k$.

\item A fraction $\Gamma X_{N+1}$ is reflected. From this fraction, an amount $\Gamma x_1$ escapes the gravitational well of the object, while the remaining amount $\Gamma\sum_{k=2}^{N+1}x_n$ is gravitationally lensed back to the object.
\item A fraction $\mathrm{T}X_{N+1}$ travels across the object without interaction.
\end{itemize}

For the $(N+2)-$interval:
\begin{itemize}
\item There is an injection of energy $x$ onto the horizonless object. 

\item A fraction of energy $(1-\Delta)\tilde{\Gamma}\sum_{k=1}^{N+1}x_k$ returns to the surface after being re-emitted in the previous interval.

\item A fraction of energy $\Gamma\sum_{k=2}^{N+1}x_k$ returns to the surface after being reflected in the previous interval.

\item The total amount of incident energy is $X_{N+2}=\sum_{k=1}^{N+2}x_k$, where $x_{k}=[\Gamma+(1-\Delta)\tilde{\Gamma}]x_{k-1}$ for $k\leq N+1$ and $x_{N+2}=[\Gamma+(1-\Delta)\tilde{\Gamma}]x_{N+1}+(1-\Delta)\tilde{\kappa}x_1$.

\item{A fraction of energy $\kappa X_{N+2}$ is permanently absorbed by the object.}

\item A fraction $\tilde{\kappa}X_{N+2}$ is temporarily absorbed by the object, and will be re-emitted after a time $N\tau_{\rm S}$.

\item A fraction $\tilde{\kappa}x_2$ is re-emitted after being temporarily absorbed by the object, while a fraction $\tilde{\Gamma}X_{N+2}$ is re-emitted instantaneously. From these fractions, an amount $\Delta(\tilde{\Gamma}X_{N+2}+\tilde{\kappa}x_2)$ escapes the gravitational well of the object, while the remaining amount $(1-\Delta)(\tilde{\Gamma}X_{N+2}+\tilde{\kappa}x_2)$ is gravitationally lensed back to the object. Let us define $\epsilon_{N+2}=\Delta(\tilde{\Gamma}x_{N+2}+\tilde{\kappa}x_2)$, so that the total energy that escapes is $E_{N+2}=\sum_{k=1}^{N+2}\epsilon_k$.
\end{itemize}

\acknowledgments 
The authors are grateful to Ramesh Narayan for valuable comments on improving this paper and for previous communications on the subject. RCR acknowledges financial support through a research grant (29405) from VILLUM fonden. FDF acknowledges financial support by Japan Society for the Promotion of Science Grants-in-Aid for international research fellow No. 21P21318. 
SL acknowledges funding from the Italian Ministry of Education and  Scientific Research (MIUR)  under the grant  PRIN MIUR 2017-MB8AEZ. 
MV was supported by the Marsden Fund, via a grant administered by the Royal Society of New Zealand. 

\bibliography{refs}

\begin{thebibliography}{34}
\expandafter\ifx\csname natexlab\endcsname\relax\def\natexlab#1{#1}\fi
\expandafter\ifx\csname bibnamefont\endcsname\relax
  \def\bibnamefont#1{#1}\fi
\expandafter\ifx\csname bibfnamefont\endcsname\relax
  \def\bibfnamefont#1{#1}\fi
\expandafter\ifx\csname citenamefont\endcsname\relax
  \def\citenamefont#1{#1}\fi
\expandafter\ifx\csname url\endcsname\relax
  \def\url#1{\texttt{#1}}\fi
\expandafter\ifx\csname urlprefix\endcsname\relax\def\urlprefix{URL }\fi
\providecommand{\bibinfo}[2]{#2}
\providecommand{\eprint}[2][]{\url{#2}}

\bibitem[{\citenamefont{Carballo-Rubio
  et~al.}(2020{\natexlab{a}})\citenamefont{Carballo-Rubio, Di~Filippo,
  Liberati, and Visser}}]{Carballo-Rubio:2019fnb}
\bibinfo{author}{\bibfnamefont{R.}~\bibnamefont{Carballo-Rubio}},
  \bibinfo{author}{\bibfnamefont{F.}~\bibnamefont{Di~Filippo}},
  \bibinfo{author}{\bibfnamefont{S.}~\bibnamefont{Liberati}}, \bibnamefont{and}
  \bibinfo{author}{\bibfnamefont{M.}~\bibnamefont{Visser}},
  \bibinfo{journal}{Phys. Rev. D} \textbf{\bibinfo{volume}{101}},
  \bibinfo{pages}{084047} (\bibinfo{year}{2020}{\natexlab{a}}),
  \eprint{1911.11200}.

\bibitem[{\citenamefont{Carballo-Rubio
  et~al.}(2020{\natexlab{b}})\citenamefont{Carballo-Rubio, Di~Filippo,
  Liberati, and Visser}}]{Carballo-Rubio:2019nel}
\bibinfo{author}{\bibfnamefont{R.}~\bibnamefont{Carballo-Rubio}},
  \bibinfo{author}{\bibfnamefont{F.}~\bibnamefont{Di~Filippo}},
  \bibinfo{author}{\bibfnamefont{S.}~\bibnamefont{Liberati}}, \bibnamefont{and}
  \bibinfo{author}{\bibfnamefont{M.}~\bibnamefont{Visser}},
  \bibinfo{journal}{Class. Quant. Grav.} \textbf{\bibinfo{volume}{37}},
  \bibinfo{pages}{14} (\bibinfo{year}{2020}{\natexlab{b}}),
  \eprint{1908.03261}.

\bibitem[{\citenamefont{Carballo-Rubio
  et~al.}(2022{\natexlab{a}})\citenamefont{Carballo-Rubio, Di~Filippo,
  Liberati, and Visser}}]{Carballo-Rubio:2022nuj}
\bibinfo{author}{\bibfnamefont{R.}~\bibnamefont{Carballo-Rubio}},
  \bibinfo{author}{\bibfnamefont{F.}~\bibnamefont{Di~Filippo}},
  \bibinfo{author}{\bibfnamefont{S.}~\bibnamefont{Liberati}}, \bibnamefont{and}
  \bibinfo{author}{\bibfnamefont{M.}~\bibnamefont{Visser}}
  (\bibinfo{year}{2022}{\natexlab{a}}), \eprint{2211.05817}.

\bibitem[{\citenamefont{Laughlin}(2003)}]{Laughlin:2003yh}
\bibinfo{author}{\bibfnamefont{R.~B.} \bibnamefont{Laughlin}},
  \bibinfo{journal}{Int. J. Mod. Phys. A} \textbf{\bibinfo{volume}{18}},
  \bibinfo{pages}{831} (\bibinfo{year}{2003}), \eprint{gr-qc/0302028}.

\bibitem[{\citenamefont{Mazur and Mottola}(2004)}]{Mazur:2004fk}
\bibinfo{author}{\bibfnamefont{P.~O.} \bibnamefont{Mazur}} \bibnamefont{and}
  \bibinfo{author}{\bibfnamefont{E.}~\bibnamefont{Mottola}},
  \bibinfo{journal}{Proc. Nat. Acad. Sci.} \textbf{\bibinfo{volume}{101}},
  \bibinfo{pages}{9545} (\bibinfo{year}{2004}), \eprint{gr-qc/0407075}.

\bibitem[{\citenamefont{Mazur and Mottola}(2023)}]{Mazur:2001fv}
\bibinfo{author}{\bibfnamefont{P.~O.} \bibnamefont{Mazur}} \bibnamefont{and}
  \bibinfo{author}{\bibfnamefont{E.}~\bibnamefont{Mottola}},
  \bibinfo{journal}{Universe} \textbf{\bibinfo{volume}{9}}, \bibinfo{pages}{88}
  (\bibinfo{year}{2023}), \eprint{gr-qc/0109035}.

\bibitem[{\citenamefont{Carballo-Rubio}(2018)}]{Carballo-Rubio:2017tlh}
\bibinfo{author}{\bibfnamefont{R.}~\bibnamefont{Carballo-Rubio}},
  \bibinfo{journal}{Phys. Rev. Lett.} \textbf{\bibinfo{volume}{120}},
  \bibinfo{pages}{061102} (\bibinfo{year}{2018}), \eprint{1706.05379}.

\bibitem[{\citenamefont{Arrechea et~al.}(2021)\citenamefont{Arrechea,
  Barcel\'o, Carballo-Rubio, and Garay}}]{Arrechea:2021pvg}
\bibinfo{author}{\bibfnamefont{J.}~\bibnamefont{Arrechea}},
  \bibinfo{author}{\bibfnamefont{C.}~\bibnamefont{Barcel\'o}},
  \bibinfo{author}{\bibfnamefont{R.}~\bibnamefont{Carballo-Rubio}},
  \bibnamefont{and} \bibinfo{author}{\bibfnamefont{L.~J.} \bibnamefont{Garay}},
  \bibinfo{journal}{Phys. Rev. D} \textbf{\bibinfo{volume}{104}},
  \bibinfo{pages}{084071} (\bibinfo{year}{2021}), \eprint{2105.11261}.

\bibitem[{\citenamefont{Arrechea et~al.}(2022)\citenamefont{Arrechea,
  Barcel\'o, Carballo-Rubio, and Garay}}]{Arrechea:2021xkp}
\bibinfo{author}{\bibfnamefont{J.}~\bibnamefont{Arrechea}},
  \bibinfo{author}{\bibfnamefont{C.}~\bibnamefont{Barcel\'o}},
  \bibinfo{author}{\bibfnamefont{R.}~\bibnamefont{Carballo-Rubio}},
  \bibnamefont{and} \bibinfo{author}{\bibfnamefont{L.~J.} \bibnamefont{Garay}},
  \bibinfo{journal}{Sci. Rep.} \textbf{\bibinfo{volume}{12}},
  \bibinfo{pages}{15958} (\bibinfo{year}{2022}), \eprint{2110.15808}.

\bibitem[{\citenamefont{Carballo-Rubio
  et~al.}(2018{\natexlab{a}})\citenamefont{Carballo-Rubio, Di~Filippo,
  Liberati, and Visser}}]{Carballo-Rubio:2018jzw}
\bibinfo{author}{\bibfnamefont{R.}~\bibnamefont{Carballo-Rubio}},
  \bibinfo{author}{\bibfnamefont{F.}~\bibnamefont{Di~Filippo}},
  \bibinfo{author}{\bibfnamefont{S.}~\bibnamefont{Liberati}}, \bibnamefont{and}
  \bibinfo{author}{\bibfnamefont{M.}~\bibnamefont{Visser}},
  \bibinfo{journal}{Phys. Rev. D} \textbf{\bibinfo{volume}{98}},
  \bibinfo{pages}{124009} (\bibinfo{year}{2018}{\natexlab{a}}),
  \eprint{1809.08238}.

\bibitem[{\citenamefont{Cardoso and Pani}(2019)}]{Cardoso:2019rvt}
\bibinfo{author}{\bibfnamefont{V.}~\bibnamefont{Cardoso}} \bibnamefont{and}
  \bibinfo{author}{\bibfnamefont{P.}~\bibnamefont{Pani}},
  \bibinfo{journal}{Living Rev. Rel.} \textbf{\bibinfo{volume}{22}},
  \bibinfo{pages}{4} (\bibinfo{year}{2019}), \eprint{1904.05363}.

\bibitem[{\citenamefont{Akiyama et~al.}(2019)}]{EventHorizonTelescope:2019dse}
\bibinfo{author}{\bibfnamefont{K.}~\bibnamefont{Akiyama}} \bibnamefont{et~al.}
  (\bibinfo{collaboration}{Event Horizon Telescope}),
  \bibinfo{journal}{Astrophys. J. Lett.} \textbf{\bibinfo{volume}{875}},
  \bibinfo{pages}{L1} (\bibinfo{year}{2019}), \eprint{1906.11238}.

\bibitem[{\citenamefont{Akiyama et~al.}(2022)}]{EventHorizonTelescope:2022xqj}
\bibinfo{author}{\bibfnamefont{K.}~\bibnamefont{Akiyama}} \bibnamefont{et~al.}
  (\bibinfo{collaboration}{Event Horizon Telescope}),
  \bibinfo{journal}{Astrophys. J. Lett.} \textbf{\bibinfo{volume}{930}},
  \bibinfo{pages}{L17} (\bibinfo{year}{2022}).

\bibitem[{\citenamefont{Broderick and Narayan}(2006)}]{Broderick:2005xa}
\bibinfo{author}{\bibfnamefont{A.~E.} \bibnamefont{Broderick}}
  \bibnamefont{and} \bibinfo{author}{\bibfnamefont{R.}~\bibnamefont{Narayan}},
  \bibinfo{journal}{Astrophys. J. Lett.} \textbf{\bibinfo{volume}{638}},
  \bibinfo{pages}{L21} (\bibinfo{year}{2006}), \eprint{astro-ph/0512211}.

\bibitem[{\citenamefont{Broderick and Narayan}(2007)}]{Broderick:2007ek}
\bibinfo{author}{\bibfnamefont{A.~E.} \bibnamefont{Broderick}}
  \bibnamefont{and} \bibinfo{author}{\bibfnamefont{R.}~\bibnamefont{Narayan}},
  \bibinfo{journal}{Class. Quant. Grav.} \textbf{\bibinfo{volume}{24}},
  \bibinfo{pages}{659} (\bibinfo{year}{2007}), \eprint{gr-qc/0701154}.

\bibitem[{\citenamefont{Narayan and McClintock}(2008)}]{Narayan:2008bv}
\bibinfo{author}{\bibfnamefont{R.}~\bibnamefont{Narayan}} \bibnamefont{and}
  \bibinfo{author}{\bibfnamefont{J.~E.} \bibnamefont{McClintock}},
  \bibinfo{journal}{New Astron. Rev.} \textbf{\bibinfo{volume}{51}},
  \bibinfo{pages}{733} (\bibinfo{year}{2008}), \eprint{0803.0322}.

\bibitem[{\citenamefont{Broderick et~al.}(2009)\citenamefont{Broderick, Loeb,
  and Narayan}}]{Broderick:2009ph}
\bibinfo{author}{\bibfnamefont{A.~E.} \bibnamefont{Broderick}},
  \bibinfo{author}{\bibfnamefont{A.}~\bibnamefont{Loeb}}, \bibnamefont{and}
  \bibinfo{author}{\bibfnamefont{R.}~\bibnamefont{Narayan}},
  \bibinfo{journal}{Astrophys. J.} \textbf{\bibinfo{volume}{701}},
  \bibinfo{pages}{1357} (\bibinfo{year}{2009}), \eprint{0903.1105}.

\bibitem[{\citenamefont{Broderick et~al.}(2015)\citenamefont{Broderick,
  Narayan, Kormendy, Perlman, Rieke, and Doeleman}}]{Broderick:2015tda}
\bibinfo{author}{\bibfnamefont{A.~E.} \bibnamefont{Broderick}},
  \bibinfo{author}{\bibfnamefont{R.}~\bibnamefont{Narayan}},
  \bibinfo{author}{\bibfnamefont{J.}~\bibnamefont{Kormendy}},
  \bibinfo{author}{\bibfnamefont{E.~S.} \bibnamefont{Perlman}},
  \bibinfo{author}{\bibfnamefont{M.~J.} \bibnamefont{Rieke}}, \bibnamefont{and}
  \bibinfo{author}{\bibfnamefont{S.~S.} \bibnamefont{Doeleman}},
  \bibinfo{journal}{Astrophys. J.} \textbf{\bibinfo{volume}{805}},
  \bibinfo{pages}{179} (\bibinfo{year}{2015}), \eprint{1503.03873}.

\bibitem[{\citenamefont{Carballo-Rubio
  et~al.}(2022{\natexlab{b}})\citenamefont{Carballo-Rubio, Di~Filippo,
  Liberati, and Visser}}]{Carballo-Rubio:2022imz}
\bibinfo{author}{\bibfnamefont{R.}~\bibnamefont{Carballo-Rubio}},
  \bibinfo{author}{\bibfnamefont{F.}~\bibnamefont{Di~Filippo}},
  \bibinfo{author}{\bibfnamefont{S.}~\bibnamefont{Liberati}}, \bibnamefont{and}
  \bibinfo{author}{\bibfnamefont{M.}~\bibnamefont{Visser}},
  \bibinfo{journal}{JCAP} \textbf{\bibinfo{volume}{08}}, \bibinfo{pages}{055}
  (\bibinfo{year}{2022}{\natexlab{b}}), \eprint{2205.13555}.

\bibitem[{\citenamefont{Lu et~al.}(2017)\citenamefont{Lu, Kumar, and
  Narayan}}]{Lu:2017vdx}
\bibinfo{author}{\bibfnamefont{W.}~\bibnamefont{Lu}},
  \bibinfo{author}{\bibfnamefont{P.}~\bibnamefont{Kumar}}, \bibnamefont{and}
  \bibinfo{author}{\bibfnamefont{R.}~\bibnamefont{Narayan}},
  \bibinfo{journal}{Mon. Not. Roy. Astron. Soc.}
  \textbf{\bibinfo{volume}{468}}, \bibinfo{pages}{910} (\bibinfo{year}{2017}),
  \eprint{1703.00023}.

\bibitem[{\citenamefont{Cardoso and
  Pani}(2017{\natexlab{a}})}]{Cardoso:2017cqb}
\bibinfo{author}{\bibfnamefont{V.}~\bibnamefont{Cardoso}} \bibnamefont{and}
  \bibinfo{author}{\bibfnamefont{P.}~\bibnamefont{Pani}},
  \bibinfo{journal}{Nature Astron.} \textbf{\bibinfo{volume}{1}},
  \bibinfo{pages}{586} (\bibinfo{year}{2017}{\natexlab{a}}),
  \eprint{1709.01525}.

\bibitem[{\citenamefont{Taylor and Starkman}(2022)}]{Taylor:2022pgs}
\bibinfo{author}{\bibfnamefont{Q.}~\bibnamefont{Taylor}} \bibnamefont{and}
  \bibinfo{author}{\bibfnamefont{G.~D.} \bibnamefont{Starkman}},
  \bibinfo{journal}{Phys. Rev. D} \textbf{\bibinfo{volume}{105}},
  \bibinfo{pages}{043526} (\bibinfo{year}{2022}), \eprint{2201.08948}.

\bibitem[{\citenamefont{Carballo-Rubio
  et~al.}(2022{\natexlab{c}})\citenamefont{Carballo-Rubio, Cardoso, and
  Younsi}}]{Carballo-Rubio:2022aed}
\bibinfo{author}{\bibfnamefont{R.}~\bibnamefont{Carballo-Rubio}},
  \bibinfo{author}{\bibfnamefont{V.}~\bibnamefont{Cardoso}}, \bibnamefont{and}
  \bibinfo{author}{\bibfnamefont{Z.}~\bibnamefont{Younsi}},
  \bibinfo{journal}{Phys. Rev. D} \textbf{\bibinfo{volume}{106}},
  \bibinfo{pages}{084038} (\bibinfo{year}{2022}{\natexlab{c}}),
  \eprint{2208.00704}.

\bibitem[{\citenamefont{Eichhorn
  et~al.}(2023{\natexlab{a}})\citenamefont{Eichhorn, Held, and
  Johannsen}}]{Eichhorn:2022oma}
\bibinfo{author}{\bibfnamefont{A.}~\bibnamefont{Eichhorn}},
  \bibinfo{author}{\bibfnamefont{A.}~\bibnamefont{Held}}, \bibnamefont{and}
  \bibinfo{author}{\bibfnamefont{P.-V.} \bibnamefont{Johannsen}},
  \bibinfo{journal}{JCAP} \textbf{\bibinfo{volume}{01}}, \bibinfo{pages}{043}
  (\bibinfo{year}{2023}{\natexlab{a}}), \eprint{2204.02429}.

\bibitem[{\citenamefont{Eichhorn
  et~al.}(2023{\natexlab{b}})\citenamefont{Eichhorn, Gold, and
  Held}}]{Eichhorn:2022fcl}
\bibinfo{author}{\bibfnamefont{A.}~\bibnamefont{Eichhorn}},
  \bibinfo{author}{\bibfnamefont{R.}~\bibnamefont{Gold}}, \bibnamefont{and}
  \bibinfo{author}{\bibfnamefont{A.}~\bibnamefont{Held}},
  \bibinfo{journal}{Astrophys. J.} \textbf{\bibinfo{volume}{950}},
  \bibinfo{pages}{117} (\bibinfo{year}{2023}{\natexlab{b}}),
  \eprint{2205.14883}.

\bibitem[{\citenamefont{Cardoso and
  Pani}(2017{\natexlab{b}})}]{Cardoso:2017njb}
\bibinfo{author}{\bibfnamefont{V.}~\bibnamefont{Cardoso}} \bibnamefont{and}
  \bibinfo{author}{\bibfnamefont{P.}~\bibnamefont{Pani}}
  (\bibinfo{year}{2017}{\natexlab{b}}), \eprint{1707.03021}.

\bibitem[{\citenamefont{Zulianello et~al.}(2021)\citenamefont{Zulianello,
  Carballo-Rubio, Liberati, and Ansoldi}}]{Zulianello:2020cmx}
\bibinfo{author}{\bibfnamefont{A.}~\bibnamefont{Zulianello}},
  \bibinfo{author}{\bibfnamefont{R.}~\bibnamefont{Carballo-Rubio}},
  \bibinfo{author}{\bibfnamefont{S.}~\bibnamefont{Liberati}}, \bibnamefont{and}
  \bibinfo{author}{\bibfnamefont{S.}~\bibnamefont{Ansoldi}},
  \bibinfo{journal}{Phys. Rev. D} \textbf{\bibinfo{volume}{103}},
  \bibinfo{pages}{064071} (\bibinfo{year}{2021}), \eprint{2005.01837}.

\bibitem[{\citenamefont{Carballo-Rubio
  et~al.}(2018{\natexlab{b}})\citenamefont{Carballo-Rubio, Kumar, and
  Lu}}]{Carballo-Rubio:2018vin}
\bibinfo{author}{\bibfnamefont{R.}~\bibnamefont{Carballo-Rubio}},
  \bibinfo{author}{\bibfnamefont{P.}~\bibnamefont{Kumar}}, \bibnamefont{and}
  \bibinfo{author}{\bibfnamefont{W.}~\bibnamefont{Lu}}, \bibinfo{journal}{Phys.
  Rev. D} \textbf{\bibinfo{volume}{97}}, \bibinfo{pages}{123012}
  (\bibinfo{year}{2018}{\natexlab{b}}), \eprint{1804.00663}.

\bibitem[{\citenamefont{Misner and Sharp}(1964)}]{Misner:1964je}
\bibinfo{author}{\bibfnamefont{C.~W.} \bibnamefont{Misner}} \bibnamefont{and}
  \bibinfo{author}{\bibfnamefont{D.~H.} \bibnamefont{Sharp}},
  \bibinfo{journal}{Phys. Rev.} \textbf{\bibinfo{volume}{136}},
  \bibinfo{pages}{B571} (\bibinfo{year}{1964}).

\bibitem[{\citenamefont{Hernandez and Misner}(1966)}]{Hernandez:1966zia}
\bibinfo{author}{\bibfnamefont{W.~C.} \bibnamefont{Hernandez}}
  \bibnamefont{and} \bibinfo{author}{\bibfnamefont{C.~W.}
  \bibnamefont{Misner}}, \bibinfo{journal}{Astrophys. J.}
  \textbf{\bibinfo{volume}{143}}, \bibinfo{pages}{452} (\bibinfo{year}{1966}).

\bibitem[{\citenamefont{Nielsen and Yeom}(2009)}]{Nielsen:2008kd}
\bibinfo{author}{\bibfnamefont{A.~B.} \bibnamefont{Nielsen}} \bibnamefont{and}
  \bibinfo{author}{\bibfnamefont{D.-h.} \bibnamefont{Yeom}},
  \bibinfo{journal}{Int. J. Mod. Phys. A} \textbf{\bibinfo{volume}{24}},
  \bibinfo{pages}{5261} (\bibinfo{year}{2009}), \eprint{0804.4435}.

\bibitem[{\citenamefont{Guendelman and Rabinowitz}(1991)}]{Guendelman:1991qb}
\bibinfo{author}{\bibfnamefont{E.~I.} \bibnamefont{Guendelman}}
  \bibnamefont{and}
  \bibinfo{author}{\bibfnamefont{A.}~\bibnamefont{Rabinowitz}},
  \bibinfo{journal}{Phys. Rev. D} \textbf{\bibinfo{volume}{44}},
  \bibinfo{pages}{3152} (\bibinfo{year}{1991}).

\bibitem[{\citenamefont{Guendelman and Rabinowitz}(1993)}]{Guendelman:1993qf}
\bibinfo{author}{\bibfnamefont{E.~I.} \bibnamefont{Guendelman}}
  \bibnamefont{and} \bibinfo{author}{\bibfnamefont{A.~I.}
  \bibnamefont{Rabinowitz}}, \bibinfo{journal}{Phys. Rev. D}
  \textbf{\bibinfo{volume}{47}}, \bibinfo{pages}{3474} (\bibinfo{year}{1993}),
  \bibinfo{note}{[Erratum: Phys.Rev.D 48, 2961 (1993)]}.

\bibitem[{\citenamefont{Brustein et~al.}(2023)\citenamefont{Brustein, Medved,
  Shindelman, and Simhon}}]{Brustein:2023cvf}
\bibinfo{author}{\bibfnamefont{R.}~\bibnamefont{Brustein}},
  \bibinfo{author}{\bibfnamefont{A.~J.~M.} \bibnamefont{Medved}},
  \bibinfo{author}{\bibfnamefont{T.}~\bibnamefont{Shindelman}},
  \bibnamefont{and} \bibinfo{author}{\bibfnamefont{T.}~\bibnamefont{Simhon}}
  (\bibinfo{year}{2023}), \eprint{2301.09712}.

\end{thebibliography}

\end{document}